\documentclass[twocolumn]{aastex631}

\def\fedd{{\rm f}_{\rm Edd}}
\def\coeff{{\mathcal C}}
\newcommand{\MSun}{{\rm M}_{\sun}}
\newcommand{\dphi}{\Delta\psi_{\rm gas}}
\usepackage[dvipsnames]{xcolor}
\usepackage{amsmath}
\usepackage{CJK}
\usepackage{tabularx}
\usepackage{array}

\begin{document}
\begin{CJK*}{UTF8}{gbsn} 

\title{Chaotic migration of LISA Extreme Mass Ratio Inspirals in a turbulent accretion disk: effect on waveform de-phasing}

\correspondingauthor{Mudit Garg}
\email{mg9113@nyu.edu}

\correspondingauthor{Lucio Mayer}
\email{lucio.mayer@uzh.ch}

\author[0000-0002-9032-9103]{Mudit Garg}
\affiliation{Center for Cosmology and Particle Physics, Physics Department, New York University, New York, NY 10003, USA}

\author[0000-0002-7078-2074]{Lucio Mayer}
\affiliation{Department of Astrophysics, University of Zurich, Winterthurerstrasse 190, CH-8057 Z\"urich, Switzerland}

\author[0000-0003-3728-8231]{Yinhao Wu (吴寅昊)}\thanks{EACOA Fellow}
\affiliation{Shanghai Astronomical Observatory, Chinese Academy of Sciences, Shanghai 200030, China}
\affiliation{National Astronomical Observatory of Japan, 2-21-1 Osawa, Mitaka, Tokyo 181-8588, Japan}

\author[0000-0003-0533-0319]{Yacine Ali-Haïmoud}
\affiliation{Center for Cosmology and Particle Physics, Physics Department, New York University, New York, NY 10003, USA}

\author[0000-0001-5466-4628]{Douglas N.C. Lin (林潮)}
\affiliation{Department of Astronomy \& Astrophysics, University of California, Santa Cruz, CA 95064, USA}
\affiliation{Institute for Advanced Studies, Tsinghua University, Beijing 100084, China}

\begin{abstract}

Gravitational wave (GW) detector LISA will observe near-coalescence extreme mass ratio inspirals (EMRIs), which typically form in galactic central accretion disks. Torques from the disk can alter the GW-driven inspiral trajectory of an embedded EMRI from the vacuum expectation, leading to potentially observable GW dephasing ($\dphi$). So far, all studies compute $\dphi$ for a thin, laminar disk, with negligible flow turbulence, whereby the disk exerts the well-understood linear torque ($T_{\rm lin}$). However, these disks must be turbulent due to magneto-rotational instability in the inner regions. Hence, we present a proof-of-concept general prescription for the turbulent torque ($T_{\rm turb}$) acting on an EMRI by modeling it as a Gaussian distribution around $T_{\rm lin}$, inspired by recent global simulations that study such torques. We compute $\dphi$ for the ``golden'' circular EMRI with total source mass $M=10^6~\MSun$ and mass ratio $q=5\times10^{-5}$ in its final four-year evolution at redshift $z=0.276$ and signal-to-noise ratio (SNR) $=50$ by varying turbulence amplitude $\coeff$ ($=1$ in the aforementioned study), maximum correlation timescale ($N_{\rm max}$), Eddington ratio $\fedd$, disk aspect ratio $h_0$, and turbo-viscous coefficient $\alpha$ in a reasonable parameters space. For $N_{\rm max}=100$ orbits, we find that for $\coeff\gtrsim{10}$, $\fedd\gtrsim0.3$, $h_0\gtrsim0.03$, and $\alpha\gtrsim0.1$, dephasings {due to $T_{\rm lin}$} are unobservable but could become detectable ($\dphi>8/$SNR) if EMRIs experience turbulent torques. Hence, this work motivates running MHD simulations of accretion disks with embedded early-inspiral LISA EMRIs over long timescales to understand the imprint of the turbulent environment on their orbital parameters and gravitational waveforms.


\end{abstract}

\section{Introduction}

The adopted Laser Interferometer Space Antenna (LISA; \citealt{Colpi2024}) {mission}, together with {other under study detectors, such as} TianQin \citep{Li2025} and Taiji \citep{Gong2021}, provides an exciting opportunity to observe milli-Hz gravitational waves (GWs) in the mid-2030s. One of the primary sources of LISA is near-coalescence extreme mass ratio inspirals (EMRIs), where a stellar-mass ($\lesssim100~\MSun$) black hole (BH) inspirals into a supermassive BH ($\gtrsim10^6~\MSun$; SMBH) as those residing at the center of massive galaxies \citep{Kormendy2013}. The sensitivity of LISA should allow us to detect EMRIs up to redshift $z\sim7$ with up to $\mathcal O(10^5)$ GW cycles. The two main formation pathways for EMRIs  are two-body relaxation and resonant scattering of stellar remnants in dense nuclear star clusters in galactic nuclei surrounding a SMBH, and in-situ star formation and migration inside accretion of SMBHs \citep{AstroWG2023}, often referred to as AGN (active galactic nuclei) disks. There are also a few EMRI candidates that manifest themselves as highly variable X-ray sources, so-called quasi-periodic eruptions (QPEs), in accretion disks \citep{Miniutti2019,ArcaSedda2021}; however, even if their putative link with EMRIs is correct, they would still be thousands of years away from merger. 

The secondary BH of an EMRI originated in an AGN disk, or even captured by the disk from the nuclear star cluster population, typically experiences inward migration due to a disk-driven gravitational torque that hardens the EMRI's orbit enough to trigger a GW-driven fast in-spiral. However, the disk can remain as an environment that could deviate the EMRI's trajectory from a vacuum expectation even in the final few years of evolution in the LISA band. This deviation manifests as a dephasing ($\dphi$) in the gravitational waveform and could be observable for a suitable parameter space. In the last 10-15 years, many studies \citep{Levin2007,Yunes2011,Kocsis2011,Barausse2014,Derdzinski2021,Zwick2022,Garg2022,Speri2023,Hegade2025a,Hegade2025b,Duque2025a,Duque2025b,Zwick2024,Garg2025,Garg2026b} have quantified this dephasing using analytical and semi-analytical techniques. However, they mostly assume a thin, laminar disk \citep{Shakura1973}, where the gas flow has negligible turbulence and the disk exerts a fairly well-understood linear torque ($T_{\rm lin}$; \citealt{Tanaka2002}) in the so-called Type-I migration regime. These studies find that $\dphi$ is observable for $z\lesssim3$.

However, it is well established that AGN disks must be highly turbulent because turbulence drives mass and angular momentum transport, which, in turn, is the engine of the accretion itself. AGN disks comprise a hot magnetized plasma that can support vigorous magneto-hydrodynamic (MHD) turbulence, typically generated and sustained by the magneto-rotational instability (MRI) \citep[see, e.g.][]{Murphy2015}. In the outer, colder part of the disk, gravitational instability (GI) is also a likely source of turbulence. It is also possible that both instabilities co-exist, resulting in a complex interplay between Maxwell and gravitational stresses, and a potentially different dynamo mechanism, which has been identified in protoplanetary disks for a few years \citep{Riols2018,Riols2019,Deng2020} and only very recently studied in AGN disks \citep{Tsung2025}. Despite differences in the physical conditions and properties of these various mechanisms to generate turbulence, they are all capable of generating large turbulent velocities. {By parameterizing the effect of turbulence using a constant turbulent $\alpha$ viscosity, corresponding to $\alpha \gtrsim{0.2}$} in the case of an AGN disk \citep{Nouri2024,Trani2025,Tsung2025}. {Because mass inflow from larger scales could happen at an effective $\alpha$ larger than the local $\alpha$.}

The effects of a turbulent flow on the dynamics and migration of compact objects in disks, including EMRIs, have not yet been explored. The reason to ignore turbulence in above GW dephasing studies is partially for simplicity but also because of the lack of (M)HD simulations that can shape our understanding of how it can alter the expected disk torques on the EMRI. Here, we make a first step in addressing the impact of turbulence on the migration torques responsible for the inspiral of the secondary BH of an EMRI into the central SMBH. We will not assume a specific physical model for turbulence, since there could be many sources. We will work under simple general assumptions that treat the effect of turbulence as a stochastic forcing term superimposed on the classic linear Lindblad torque \citep{Paardekooper2010}. 

In this paper we will not address the effect of stochastic migration { in detail} on the dynamical properties and abundance of potential EMRI candidates in AGN disks prior to their entry into the LISA band, rather we will focus on its impact once the stellar-mass BHs have already entered the regime of strong GW-dominated inspiral trajectory. Hence, the emergence of a stochastic migration regime would not prevent the stellar-mass BHs from migrating inward. Therefore, in the LISA band, the main interest is to evaluate the impact of chaotic migration on the dephasing of the waveform that is conventionally attributed to the effect of gravitational torques exerted by the disk on the secondary BH. This is the main scope of our paper.

The paper is organized as follows. First, we provide the motivation behind our modeling of turbulent torque for EMRIs in \S~\ref{Sec:Inspiration}. Then, \S~\ref{Sec:Model} presents our mathematical framework for quantifying the effect of turbulence on the EMRI's inspiral. For the ``golden" EMRI, we compute and show our results in \S~\ref{Sec:Results}. We discuss our findings in \S~\ref{Sec:Discussion} and conclude with \S~\ref{Sec:Conclusion}.

\section{Inspiration for our turbulent torque model}\label{Sec:Inspiration}

\begin{figure}
    \centering
    \includegraphics[width=\linewidth]{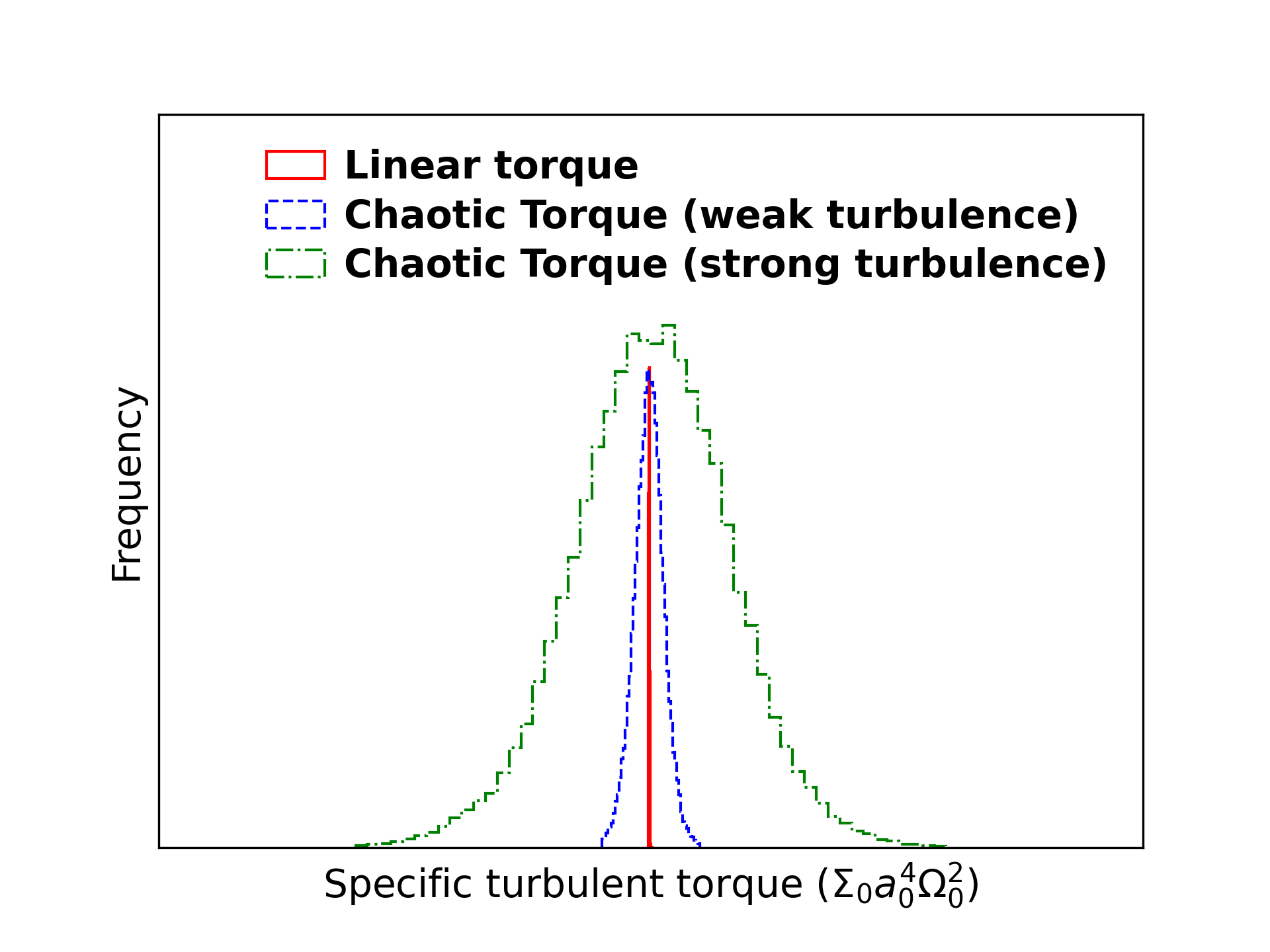}
    \caption{Qualitative distribution of the specific gas torque acting on the embedded companion in accretion disk. The red solid line represents the baseline where the turbulent perturbation is negligible, corresponding to a system governed solely by the deterministic linear torque ($T_{\rm lin}$). The blue dashed and green dot-dashed histograms illustrate the distributions of the chaotic torque under weak and strong turbulence regimes, respectively. Unlike the constant nature of the linear torque, the turbulent torque exhibits highly stochastic behavior characterized by a well-defined Gaussian dispersion. As the turbulence amplitude increases, the standard deviation of the fluctuations broadens significantly. These stochastic fluctuations drive the random-walk orbital evolution that prevents EMRIs from stalling in migration traps. Data and distribution features are modeled after the global HD simulations of \citet{Wu2024}.
    }
    \label{fig:Gaussian}
\end{figure}

The turbulent torque ($T_{\rm turb}$) prescription we employ is based on the model developed by \citet{Wu2024}, which studied the chaotic migration of low-mass planets in turbulent protoplanetary disks \citep{Chen2025}. {As demonstrated by \cite{Wu2024} (see their Figure 1), unlike the deterministic linear behavior of the Type I torque under running time-averaged conditions, the chaotic torque in a strongly turbulent environment exhibits highly stochastic behavior. This stochastic nature can be roughly described by a Gaussian distribution, which is dominated by the fluctuating component with a typical dispersion (as shown in Fig. \ref{fig:Gaussian}). Furthermore, in an earlier study utilizing the same turbulence prescription, \cite{Baruteau2010} derived the scaling relationship between the dimensionless parameter ``gamma", which characterizes the turbulence amplitude, and the equivalent turbo-viscous parameter $\alpha$. By experimenting with different initial disk aspect ratios ($h_0$), they demonstrated that the error in this relationship does not exceed 20\%.}

{Thus,} we model $T_{\rm turb}$ as a Gaussian around $T_{\rm lin}$ with variance that depends upon the binary-disk parameters as well as the level of the turbulence in the gas flow. The model is in basic agreement with the outcomes of migration in complex, nonlinear dynamical evolution regimes, such as 3D gravito-turbulent magnetized disks with multiple planets having a wide spectrum of masses \citep{Kubli2026}, and 3D MHD-disk simulations with magnetospheric accretion \citep{CevallosSoto-Zhu2026}. \citet{Wu2024} found that the turbulent torques dominate the evolution of the orbit once the mass ratio between the planet and the star falls below $\sim10^{-5}$, which corresponds to the EMRI regime in the case of AGN disks. The standard deviation of $T_{\rm turb}$ reported by \citet{Wu2024} is larger than what \citet{Nelson2005} suggest. This is most likely because the former evolves the system for $2000$ orbits, while the latter only simulate $\sim100$ orbits.

Specifically, incorporating insights from global HD simulations \citep{Wu2024}, this chaotic migration model exhibits several defining features. First, it demonstrates that in a strongly turbulent environment, the classical circum-companion density waves and flow patterns are completely disrupted. Consequently, the time-averaged residual migration torque does not converge to classical Type I expectations, but rather decays to negligible levels. { For our GW-driven EMRI system, it will never be in a steady state and will not have extremely low gas torques. Hence, we only utilize the results from the first few hundred orbits.} Second, the turbulent torque exhibits an approximately Gaussian distribution. This high degree of stochasticity shifts the orbital evolution of low-mass companions from monotonic inward migration to a slow, random-walk dominated diffusion. Third, the model indicates that sufficiently massive companions capable of marginal gap-opening are significantly less susceptible to these chaotic effects. Macroscopically, this chaotic migration mechanism suggests that stochastic torques can effectively prevent stellar-mass BHs from monotonically settling and crowding into universal ``migration traps" \citep{Horn2012,Bellovary2016,Secunda2020,Grishin2024,McKernan2025,McPike2026} within the accretion disk, thereby theoretically reducing the expected frequency of three-body interactions and hierarchical mergers. The diminishing role of migration traps implies that a larger fraction of stellar-mass BHs should become EMRIs. 

Note that the proposition of ``migration traps" is largely inferred from numerical simulation of laminar disk flows with a single, non-accreting perturber on a circular and coplanar orbit. Concurrent accretion onto migrating EMRIs can introduce azimuthal asymmetries in the horseshoe orbits and modify the torque magnitude and direction \citep{lichenlin2024, pan2026, ida2026}. In addition to turbulence, density wakes from nearby perturbers can prevent the formation of the gap \citep[see, e.g.][]{Bryden2000,Li2022}.

\section{Modeling chaotic migration of EMRIs}\label{Sec:Model}

Let us consider a circular EMRI system with a total redshifted mass $M_z$, a secondary-to-primary mass ratio $q\lesssim10^{-4}$, and a semi-major axis (SMA) $a$ embedded in a {locally isothermal (i.e., a fixed $h_0$}) thin accretion disk. The orbit of the binary system is mainly shrinking via GW radiation with frequency
\begin{equation}
    f=\frac{1}{\pi}\left(\frac{GM_z}{a^3}\right),
\end{equation}
which is twice the orbital frequency $\Omega/2\pi$ and the rate of change of SMA at the leading order \citep{Peters1964}:
\begin{equation}
    \dot{a}_{\rm GW}=-\frac{64}{5}\frac{G^3}{c^5}\frac{q M_z^3}{a^3},
\end{equation}
where we have approximated the symmetric mass ratio $q/(1+q)^2\approx q$ because $q\ll1$.\\

{In this work, we employ the radiation-dominated inner surface-density disk profile,\footnote{See Appendix~\ref{App:Sigma_vis} for results for the viscous disk.} which should be more applicable for the LISA EMRIs since they will have at most $\sim10~r_s$ separations. For a given $\alpha$ and Eddington ratio $\fedd$,\footnote{$\fedd$ is the ratio of the accretion rate $\dot M$ onto the EMRI and the Eddington accretion rate $\dot M_{\rm Edd}=M_z/50$ Myr for our assumed $10\%$ radiative efficiency.} $\Sigma$ profile is \citep{Grishin2024}
\begin{align}\label{Eq:Sigma}
    \Sigma&=16.5\left(\frac{a}{r_{\rm s}}\right)^{\frac32}\left(\frac{\fedd}{1.0}\right)^{-1}\left(\frac{\alpha}{0.1}\right)^{-1}{\rm g}\cdot{\rm cm}^{-2},
\end{align}
where $r_s\equiv2GM_z/c^2$ is the Schwarzschild radius. 
Because EMRIs that merge in the LISA band will have at most $\sim10~r_s$ separation and the dephasing computed later accumulates mostly at earlier SMAs ; Hence, we will assume constant surface density ($\Sigma_0$) defined at the initial SMA $a_0$ throughout their evolution \citep[see, e.g.][]{Garg2022}. This is only reasonable if the radial power-law of $\Sigma$ is not too steep (i.e. $\lesssim2$). Lastly, since dephasings studied later accumulate mostly at the initial inspiral, our later computed results will not be affected by the `upper hook shape' that the above surface density profile will experience near the innermost stable circular orbit (ISCO) separation of $3~r_s$.}

In a laminar thin disk, where gas turbulence is negligible, an {
isolated} EMRI {with a circular orbit} will experience the linear torque ($T_{\rm lin}$){, which can be expressed for our disk profile in Eq.~\ref{Eq:Sigma} as \citep{Tanaka2002,Jimenez2017} }
\begin{align}
    T_{\rm{lin}}&\approx-{1.25}(q/h_0)^2\Sigma_0 a^4\Omega^2.
\end{align}
{ This quantity includes both differential the Lindblad torque and the unsaturated corotation torque due to viscous diffusion along the horseshoe
streamlines \citep{Paardekooper2010, paardekooper2011}. The pre-factor $-1.25$ comes from considering radial power-law dependence of surface density (i.e., $3/2$) together with $1/2$ radial power-law dependence for the mid-plane temperature under our assumption of a locally isothermal disk and $1.4$ adiabatic index similar to \citet{Jimenez2017}. For the parameter space we explore later, the above pre-factor does not change by more than $\sim1\%$.}

{However, in a highly turbulent gas flow, the dominant 
contribution from the corotation torque is strongly affected by the 
local intrinsic stochastic flow. The instantaneous total (Lindblad plus corotation) torque fluctuates in sign as well as magnitude, and
thereby suppresses the global radial dependence on the $\Sigma$ distribution. Averaged over time,} the disk exerts a turbulent torque ($T_{\rm{turb}}$) on the binary, which can be modeled as a Gaussian distribution (represented by $\mathcal{N}$) around $T_{\rm lin}$ based upon the recent global HD study \citep{Wu2024,Baruteau2010}:
\begin{align}\label{Eq:Tturb}
    T_{\rm{turb}}&\approx \mathcal{N}(T_{\rm lin},\sigma^2_{T_{\rm turb}}),\\
    {\rm where}~\sigma_{T_{\rm turb}}&={0.29\coeff q\left(\frac{\fedd}{1.0}\right)\left(\frac{h_0}{0.03}\right)\left(\frac{\alpha}{0.1}\right)^{\frac32}}\Sigma_0  a^4\Omega^2\nonumber.
\end{align}
{Here we have explicitly expressed the dimensionless parameter ``gamma" in \citet{Wu2024} and introduce a new dimensionless turbulence amplitude $\coeff$, which is defined relative to the dynamically relevant turbulent torque threshold in \citet{Wu2024} simulations. Therefore, $\coeff=1$ corresponds to our underlying HD simulations.
It is to be noted that $\sigma_{T_{\rm turb}}\propto\fedd^0\alpha^{1/2}h_0$ in \citet{Wu2024} for constant $\Sigma_0=1$ at $a=a_0$, irrespective of other parameters. Hence, we also preserve the same scalings for $\sigma_{T_{\rm turb}}$ here by setting power-law dependence on $\{\fedd,\alpha,h_0\}$ to be $\{1,3/2,1\}$, respectively, for our physical $\Sigma\propto\fedd^{-1}\alpha^{-1}$ in Eq.~\eqref{Eq:Sigma}. This is also physically sensible, since higher $\alpha$ should lead to stronger turbulence.}

{In Fig.~\ref{fig:Gaussian}, the weak turbulence corresponds to $\coeff=0.1$ and the strong one results from $\coeff=1$. Thus,} a rule of thumb is that {$\coeff\lesssim0.1$} implies a laminar flow and {$\coeff\gtrsim0.1$} results in a turbulent flow. Hence, we are assuming that a highly turbulent disk induces fluctuations of order $\sigma_{T_{\rm Turb}}$ around the linear torque. The functional form for $\sigma_{T_{\rm Turb}}$ has been calibrated against simulations with different $q$ and {``gamma"} in \citet{Wu2024}. 

One can { then} agnostically express the gas torque onto an EMRI by {introducing dimensionless parameters $\xi$, $\mu_{\xi}$, and $\sigma_{\xi}$ as follows
\begin{align}\label{Eq:Tgas}
    T_{\rm gas}&=\xi {\Sigma_0}a^4\Omega^2,\\
    {\rm where}~\xi&=\begin{cases}
    \mu_\xi, {~\rm e.g.}~\coeff\lesssim0.1~(T_{\rm gas}=T_{\rm lin}),\\
    \mathcal{N}\left(\mu_\xi,\sigma^2_\xi\right),{~\rm e.g.}~\coeff\gtrsim0.1~(T_{\rm gas}=T_{\rm turb}),
    \end{cases}\nonumber\\
    {\rm with}~\mu_\xi&=-1.25(q/h_0)^2~{\rm and}\nonumber\\
    \sigma_{\xi}&=0.29\coeff q\left(\frac{\fedd}{1.0}\right)\left(\frac{h_0}{0.03}\right)\left(\frac{\alpha}{0.1}\right)^{\frac32}.\nonumber
\end{align}
Here $\mu_{\xi}$ is the normalization of the linear torque and $\sigma_{\xi}$ is the Gaussian width of the normalized turbulent torque.}

Performing a simple Newtonian torque balance gives us gas-driven rate of change of SMA due to $T_{\rm gas}$ \citep{Garg2024b}:
\begin{align}
     \dot{a}_{\rm gas}&=\frac{2 T_{\rm gas}}{q M a\Omega}={\mathcal{A}}\left(\frac{a}{GM_z/c^2}\right)^{4.5}\dot{a}_{\rm GW},\\
     {\rm where}~\mathcal{A}&=-\frac{5\xi}{32}{\Sigma_0}\frac{G^2M_z}{c^4q^2}.\nonumber
\end{align}

Therefore, the cumulative gas-induced dephasing in the {GW stationary phase approximation\footnote{Since gas does not non-negligibly changes a given EMRI's mass during the four-year observations, even if it accretes at the Eddington limit, the corresponding GW amplitude is unaffected, and only the phase has potentially observable gas-induced dephasing in the LISA band.} phase \citep{Cutler1994}} from a SMA $a$ to the coalescence \citep{Garg2024b} for a given $\xi$ is
\begin{equation}\label{Eq:Dephasing}
    \Delta\psi_{\rm gas}(a,\xi)=-\psi^{(0)}_{\rm TF2}\frac{{20\mathcal{A}}}{119}v^{-9},
\end{equation}
where $v\equiv(GM_z\pi f/c^3)^{\frac13}$ is the ratio of the orbital velocity to $c$ and the perturbation parameter for the post-Newtonian expansion, and $\psi^{(0)}_{\rm TF2}\equiv-(3/128q)v^{-5}$ is the leading-order phase term (denoted by the $(0)$ superscript) for the TaylorF2 circular binary frequency-domain template \citep{Buonanno2009}. At the first order, a dephasing is observable if its more than ${\Delta \psi_{\rm crit}\equiv}~8/$SNR \citep{Kocsis2011,Garg2022}, where SNR is the signal-to-noise ratio of the source in the LISA band.

In the turbulent regime (${\coeff\gtrsim0.1}$), Figure~1 of \citet{Wu2024} suggests that the corresponding $\xi$ changes after a few to several tens of binary orbits. {In our framework, the orbital evolution is driven by a stochastic random walk. The amplitude of each step in this random walk is governed by the Gaussian standard deviation. As explicitly demonstrated by the torque distribution histograms in Fig.~\ref{fig:Gaussian}, the chaotic torque follows a Gaussian dispersion whose width depends directly on the turbulence amplitude. Crucially, the time interval between these steps, meaning how often the stochastic forcing updates, which is dictated by the correlation timescale (or eddy turnover time) of the turbulent structures. Because larger turbulent structures naturally have longer physical lifetimes than smaller ones, this yields a characteristic correlation timescale ranging from a few to several tens of binary orbits. It is over this specific timescale that a macroscopic density asymmetry, and its associated gravitational torque, persists before dissipating into a newly generated random configuration.}

Hence, agnostically we can model this by selecting a new $\xi$ after $N$ GW-driven inspiral orbits, where $N$ is chosen from a uniform distribution between $1$ to $N_{\rm max}$ {(i.e., the maximum correlation timescale)}:
\begin{equation}\label{Eq:uniform}
    N\sim\mathcal{U}(1,N_{\rm max}).
\end{equation}
Therefore, a gas disk will exert a turbulent gas torque with a different sign and relative magnitude to $T_{\rm lin}$ on different spans of the GW-driven inspiral of the embedded EMRI that spends a total of $N_{\rm LISA}$ orbits during its final four years in the LISA band. Suppose there are $K$ such intervals then there will be $K$ tuples of $(N_j,\xi_j,a_{j})$, such that
\begin{equation}
    \Sigma_{j=0}^{K}N_j=N_{\rm LISA},
\end{equation}
and $a_{j}$ is the initial SMA of the given $j^{\rm th}$ interval and $a_{K+1}$ is the final SMA $a_f$, which is at minimum set to the ISCO. Hence, the cumulative gas-induced dephasing for an EMRI evolving in a turbulent flow is
\begin{align}\label{Eq:DephasingTotal}
    \Delta\psi_{\rm turb}&=\Sigma_{j=0}^{K}\left(\Delta\psi_{\rm gas}(a_{j},\xi_{j})-\Delta\psi_{\rm gas}(a_{j+1},\xi_j)\right),\\
    &=\Delta_0~\Sigma^k_{j=0}\xi_j\left(a^7_{j}-a^7_{j+1}\right).\nonumber,\\
    {\rm where}~\Delta_0 &= \frac{75}{952}{\Sigma_0}\frac{G^2M_z}{c^4q^3} r_s^{-7}.\nonumber
\end{align}

Given enough realizations, $\Delta\psi_{\rm turb}$ also follows a Gaussian distribution
\begin{equation}\label{Eq:dphi_Gauss}
    \Delta\psi_{\rm turb}\sim\mathcal{N}(\Delta\psi_{\rm lin},\sigma_{\rm turb}^2),
\end{equation}
where we show in Appendix~\ref{App:A} that the mean of $\Delta\psi_{\rm turb}$ is simply $\Delta\psi_{\rm lin}$, the gas-inducing dephasing due to the linear torque. We also analytically estimate (denoted with hat) the variance of the turbulent dephasing (see Appendix~\ref{App:A} for details) and express it together with $\Delta\psi_{\rm lin}$ as
\begin{align}\label{Eq:analytical}
    \Delta\psi_{\rm lin}&=\Delta_0~\mu_\xi(a^7_0-a_f^7),\\
    \hat\sigma_{\rm turb}&=\frac{5}{17} \sqrt{\frac{5\pi}{3\sqrt{2}}} \frac{G^2M_z}{c^4q^{\frac52}}\left(\frac{a_0}{r_s}\right)^{\frac{23}{4}}{\Sigma_0} \sigma_\xi \sqrt{N_{\rm max}}.\nonumber
\end{align}

In the next section, we compute results for a particular system.

\section{Results}\label{Sec:Results}

Now we consider the ``golden" LISA EMRI \citep{Speri2023} in the final four years of evolution in the LISA band with parameters
\begin{align}\label{Eq:golden}
    M_z&=10^6(1+0.276)~\MSun,\\
    q&=5\times 10^{-5},\nonumber\\
    z&=0.276,\nonumber\\
    a_0&=7.74~r_s,\nonumber\\
    a_f&=3~r_s,\nonumber\\
    {\rm SNR}&=50,\nonumber\\
    {\Delta\psi_{\rm crit}}&{=0.16~{\rm rad}},\nonumber\\
    N_{\rm LISA}&\approx 7\times10^4.\nonumber
\end{align}

If we plug these numbers into Eq.~\eqref{Eq:analytical}, then we get these analytical expectations for the Gaussian distribution of the turbulent dephasing for two disk profiles in Eq.~\eqref{Eq:Sigma}:
{
\begin{align}\label{Eq:analytical_golden}
    \frac{\Delta\psi^{\rm(golden)}_{\rm lin}}{\rm rad}&=-1.81\times 10^{-2}\left(\frac{\fedd}{1.0}\right)^{-1}\left(\frac{\alpha}{0.1}\right)^{-1}\left(\frac{h_0}{0.03}\right)^{-{2}},\\
    \frac{\hat\sigma^{\rm(golden)}_{\rm turb}}{\rm rad}&={2.98}\times10^{-3}\left(\frac{\alpha}{0.1}\right)^{\frac12}\left(\frac{h_0}{0.03}\right)\left(\frac{\coeff}{1.0}\right)\left(\frac{N_{\rm max}}{100}\right)^{\frac12}\nonumber.
\end{align}
}
Then we vary these five parameters of interest for $N_{\rm max}=100$, which is chosen based upon how often the torque changes magnitude or sign in \citet{Wu2024}:
\begin{align}\label{Eq:params}
    \coeff&\in \{1,10,100\},\\
    \fedd&\in \{10^{-1},10^{-1.5},1\}\nonumber\\
    \alpha&\in[10^{-1.25},10^{-0.25}]\nonumber\\
    h_0&\in[10^{-2},10^{-1}],\nonumber
\end{align}
where we vary $\alpha$ and $h_0$ over a log-uniform grid of hundred size. {We consider the above range of $\coeff$, since we find that for any $\coeff<1$, $\hat\sigma_{\rm turb}$ becomes extremely small and does not enhance the dephasing observability.}

\begin{figure*}
    \centering
    \includegraphics[width=0.9\linewidth]{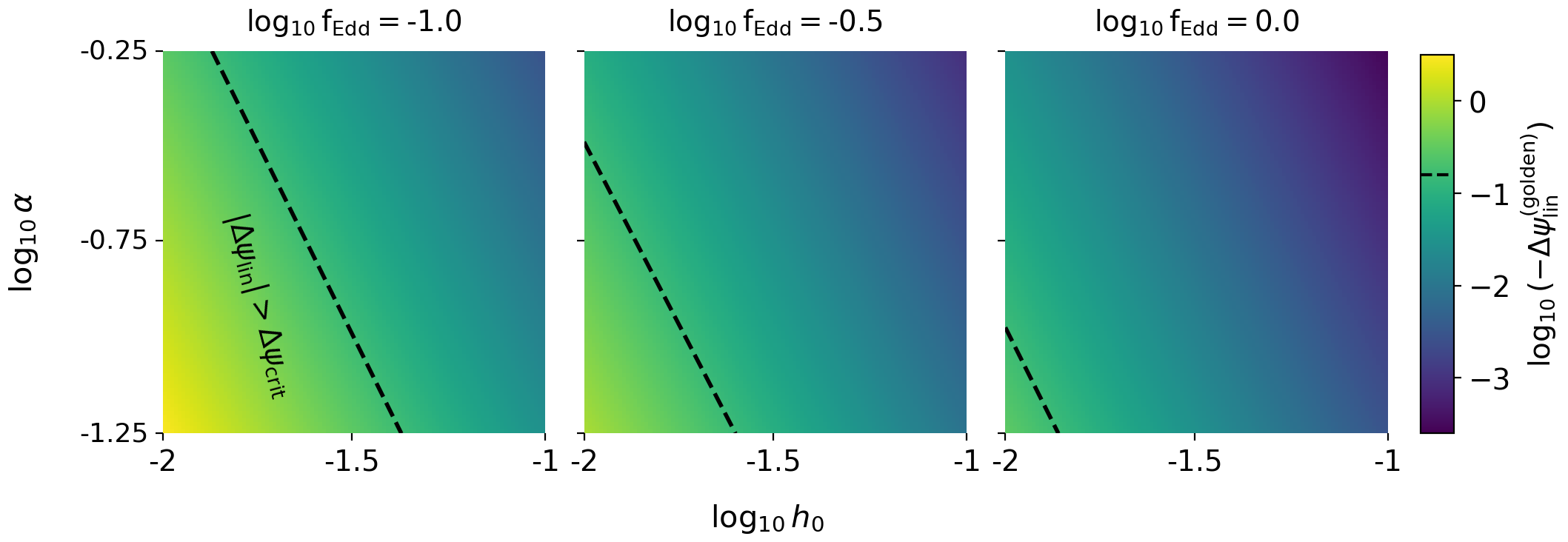}
    \caption{We show $\Delta\psi^{\rm (golden)}_{\rm lin}$, the gas-induced dephasing due to the linear torque {for the ``golden'' EMRI in Eq.~\eqref{Eq:golden}}, as a function of Eddington ratio ($\fedd$), the turbo-viscous coefficient ($\alpha$), and the disk aspect ratio ($h_0$) {for our radiation-dominated disk profile}. For each panel, we increase $\alpha$ vertically from $10^{-1.25}$ to $10^{-0.25}$ and $h_0$ horizontally from $0.01$ to $0.1$, while keeping the Eddington ratio constant. The colorbar shows the range of $\Delta\psi_{\rm lin}$ from $-10^{0.5}$ to $-10^{-3.5}$ rad. For each panel, cells left of the {dashed} black line have observable linear dephasing {due to} $\Delta\psi_{\rm lin}>{\Delta\psi_{\rm crit}=0.16~{\rm rad}}$. {Similarly, we mark this threshold in the colorbar}. Since $\Delta\psi_{\rm lin}\propto$ ${\rm f}_{\rm Edd}^{-1}\alpha^{-1}h_0^{{-2}}$ in Eq.~\eqref{Eq:analytical_golden}, the linear dephasing increases with lower $\fedd$, $\alpha$, and $h_0$.
    }
    \label{fig:Lin}
\end{figure*}

Considering the above parameter space {in Eq.~\eqref{Eq:params}}, we first show the linear dephasing $\Delta\psi^{\rm(golden)}_{\rm lin}$ in Fig.~\ref{fig:Lin} {for our golden EMRI} as the function of $\{\fedd,\alpha,h_0\}$ by using Eq.~\eqref{Eq:analytical_golden}. $\Delta\psi^{\rm(golden)}_{\rm lin}$ ranges from $-10^{0.5}$ to $-10^{-3.5}$ rad. The linear dephasings increase with lower {$\fedd$}, $\alpha$, and $h_0$. We also indicate the parameter space where linear dephasings are observable (left of {dashed} black line) whenever $\Delta\psi_{\rm lin}>{\Delta\psi_{\rm crit}}$.

Next, we consider the gas-induced dephasing due to the turbulent torque for each system of interest. We analytically estimate $\Delta\psi_{\rm turb}$ using Eq.~\eqref{Eq:analytical_golden}. We also numerically compute $\Delta\psi_{\rm turb}$ with $20000$ realizations for each combination of $\{\coeff,\fedd,\alpha,h_0\}$ {in Eq.\eqref{Eq:params}} and find that $\hat\sigma^{\rm (golden)}_{\rm turb}$ is only $\sim10\%$ more than the numerical result across the parameter space as shown in Fig.~\ref{fig:Sigma_error}. The error is mostly coming from suppressing late-inspiral dephasing with respect to the early-inspiral evolution while deriving $\hat\sigma_{\rm turb}$ in Appendix~\ref{App:A}.

To narrow down the parameter space where gas-induced dephasing could become observable due to the turbulent torque but remains unobservable due to the linear torque, we define two new quantities
\begin{align}\label{Eq:Lambda}
    \Lambda_{\rm lin}&={\log_{10}}\left(\frac{|\Delta\psi_{\rm lin}|}{{\Delta\psi_{\rm crit}}}\right),\\
     \Lambda_{\rm turb}&={\log_{10}}\left(\frac{|\Delta\psi_{\rm lin}|+2\hat\sigma_{\rm turb}}{{\Delta\psi_{\rm crit}}}\right).\nonumber
\end{align}
{Since LISA will only observe one realization out of the different possible values of the turbulent dephasing's Gaussian distribution $\Delta\psi_{\rm turb}$ in Eq.~\eqref{Eq:dphi_Gauss}, with the most likely values being its mean $\Delta\psi_{\rm lin}$, { thus} $|\Delta\psi_{\rm lin}|+2\hat\sigma_{\rm turb}$ represents the highest absolute value that LISA can realistically observe. Therefore, computing results for $\Lambda_{\rm turb}$ and comparing them with $\Lambda_{\rm lin}$ can provide an upper bound on the extent to which turbulence can enhance gas-induced GW dephasing.}

\begin{figure*}
    \centering
    \includegraphics[width=\linewidth]{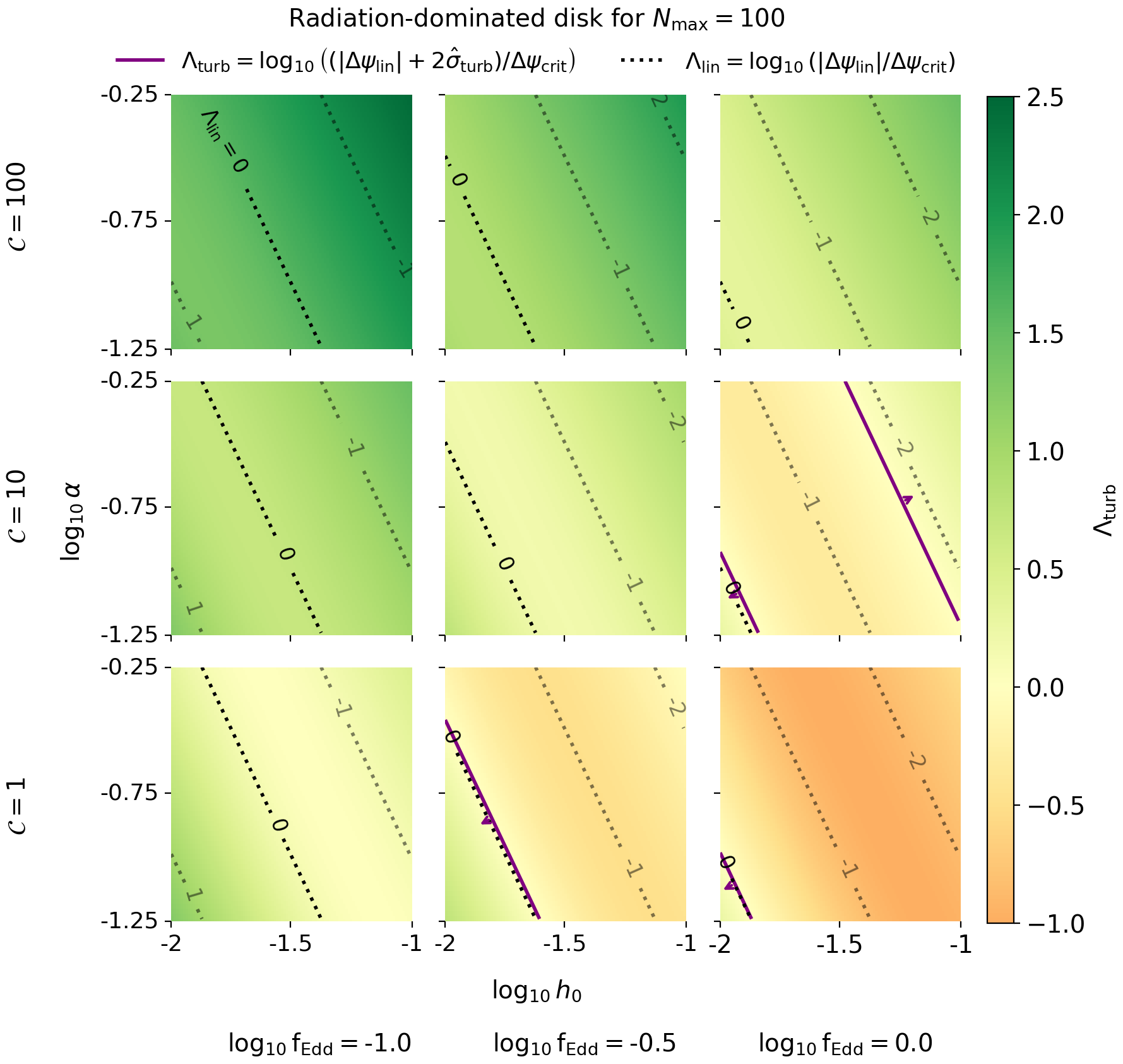}
    \caption{For the radiation-dominated disk-embedded ``golden'' EMRI in Eq.~\eqref{Eq:golden}, we show (see Eq.~\eqref{Eq:Lambda}) $\Lambda_{\rm turb}$ ($=0$; solid purple lines) as a function of $\{\coeff,\fedd,\alpha,h_0\}$ and overlay $\Lambda_{\rm lin}$ contour lines from Fig.~\ref{fig:Lin} (dotted black lines with $\Lambda_{\rm lin}=0$ shown prominently). Both $\Lambda_{\rm turb}$ and $\Lambda_{\rm lin}$ are defined relative to the minimum LISA-detectable dephasing $\Delta\psi_{\rm crit}\equiv8/{\rm SNR}=0.16$ rad, and signify whether the maximal turbulent dephasing $|\Delta\psi_{\rm lin}|+2\hat\sigma_{\rm turb}$ and the linear dephasing are LISA-observable, respectively. We use little arrows to point out the direction, where $\Lambda_{\rm turb}>0$ due to the opposing behaviors of $\Delta\psi_{\rm lin}\propto\fedd^{-1}\alpha^{-1}h_0^{-2}$ and $\hat\sigma_{\rm turb}\propto\alpha^{1/2}h_0$ in different parameter regions. Each row is same as Fig.~\ref{fig:Lin}, and we vertically increase  $\coeff$ from {$1$ to $100$} row-by-row. The colorbar is centered at $0$ and ranges from $-1$ to $2.5$.
    These results illustrate that for an extremely high turbulence gas flow ($\coeff\gtrsim100$; top row), LISA EMRIs could have detectable gas-induced dephasing across our parameter spaces. Otherwise, either for $\coeff=10$ and $\fedd\gtrsim0.03$, regions of parameter space emerge for $h_0\gtrsim0.03$ and $\alpha\gtrsim0.1$, where the maximal turbulent dephasing is observable, while the $\Delta\psi_{\rm lin}$ is not.}
    \label{Fig:Mup2Sig_rad}
\end{figure*}

{For the golden EMRI and the parameter space in Eq.~\ref{Eq:params}, we show $\Lambda_{\rm turb}$ in Fig.~\ref{Fig:Mup2Sig_rad}. For $\coeff\gtrsim100$, the whole parameter space become observable. While for $\coeff=10$ and $\fedd\gtrsim0.03$, only partial parameter space that is constrained by $h_0\gtrsim0.03$ and $\alpha\gtrsim0.1$ could become LISA-observable due to the turbulence gas flow, which is currently thought to be undetectable for a laminar gas disk.\footnote{We provide definitions of the different main variables in the main text in Table~\ref{table:variables}}.  Lastly, the maximal realistic turbulent dephasing for our parameter space is $294.46$ rad.}

\section{Discussion}\label{Sec:Discussion}
For the first time, we estimate the effect of turbulence on the orbital evolution of EMRIs in their final four years before merger in the LISA band. The inspiration for our turbulent torque model comes from a set of global HD simulations presented in \citet{Wu2024}, which provided an estimate of how the level of turbulence driven by GI changes the nature of torque when studying a planet in a proto-planetary disk with a planet-to-star mass ratio $\lesssim10^{-5}$. They find that concurrent with the linear torque, there is a stochastic Gaussian torque component that chaotically moves the planet in a random walk, such that it will not get stuck in a so-called migration trap. The same results should apply to EMRIs, as they also have $q\lesssim10^{-5}$. In the absence of the migration trap, the EMRI rates will also increase at the expense of the stellar-BH mergers. While LISA EMRIs will likely be most affected by MRI turbulence, we modeled the turbulent torque as a Gaussian around the linear torque in Eq.~\eqref{Eq:Tturb} based on fits provided by \citet{Wu2024} and introduced parameter $\coeff$ {and $N_{\rm max}$} to enhance or reduce the effect of turbulence to capture the strength of MRI with respect to GI{, and set the maximum correlation timescale, respectively.} 

We considered the ``golden" EMRI, whose parameter are given in Eq.~\eqref{Eq:golden} by varying $\coeff$, $\fedd$, $h_0$, and $\alpha$ for $N_{\rm max}=100$ in a reasonable parameters space given in Eq.~\eqref{Eq:params}. We compute our results in \S~\ref{Sec:Results} both analytically with some reasonable assumptions in Eq.~\eqref{Eq:analytical_golden} and numerically in Fig.~\ref{Fig:Mup2Sig_rad}. We find only $10\%$ error {between} our analytical approximation and the numerical results as illustrated in Fig.~\ref{fig:Sigma_error}. The turbulence can episodically increase the gas torque amplitude, potentially leading to even higher dephasing $\dphi$ and allowing us to probe gas physics around EMRIs with LISA. The region of parameter space where $\dphi$ could only become observable due to the turbulent torque is {either for $\coeff\gtrsim100$, where other parameters are less important, or for $\coeff=10$, where we need to have} $\fedd\gtrsim0.3$, $h_0\gtrsim0.03$, and $\alpha\gtrsim0.1$, as shown in Fig.~\ref{Fig:Mup2Sig_rad}; Hence, increasing the number of systems that can leave an observable imprint of the gas perturbation in the LISA band. The result computed for LISA should be worse for the TianQin detector due to the fact that its sensitivity is skewed towards higher frequency, resulting in fewer GW cycles, and should remain the same for the Taiji mission due to its overlapping frequency band.

Practically, if we indeed measure gas-induced dephasing $\dphi$ for EMRIs using LISA data, then due to the degeneracy between $\{\coeff,{N_{\rm max}}~,~\fedd,~h_0,~\alpha\}$ parameters, it can be difficult to say if its due to the turbulence torque with one set of parameters or the linear torque with different set of parameters. While for eccentric systems, one can potentially break some of these degeneracies \citep{Duque2025b} using GWs alone, for circular systems considered here, we either need improved constraints from numerical simulations \citep[see, e.g.][]{Derdzinski2025} or electromagnetic counterparts. Moreover, if the turbulent dephasing is not properly understood then it can manifest as false violations of general relativity \citep{Garg2024d}.

The stochastic torque $T_{\rm turb}$ considered here is different in nature than sub-orbital fluctuations considered in \citet{Zwick2022,Zwick2024}, which are inspired by numerical simulations of a thin, laminar, non-magnetized, and non-self-gravitating disk \citep{Derdzinski2021}. Since $T_{\rm turb}$ varies on a multiple of orbital timescale, it inherently has a different characteristic timescale than sub-orbital variations, which have a smaller characteristic timescale. Indeed, the modes assumed by \citep{Wu2024} are shearing waves with a spectrum of wavelengths, which by construction propagate on a timescale commensurate with the orbital time. Moreover, while $T_{\rm turb}$ remains different from $T_{\rm lin}$ after orbit-averaging, sub-orbital fluctuations vanish over an orbital period. However, the latter can still lead to a small dephasing \citep{Zwick2022}. This said, sub-orbital fluctuations due to small-scale turbulent eddies can be present in the MRI regime, given its broad spectrum of modes. We discuss this further in the next section.

\subsection{Caveats}

There are several caveats in this work. Even though MRI is presumably the main driver of turbulence in the inner highly ionized part of the AGN disk where  EMRIs reside during in-spiral, our description of the turbulent torque is based on \citet{Wu2024}, which was inspired by turbulence due to GI in protoplanetary disks. Yet \citet{Wu2024} considered a spectrum of 50 different modes with different wavelengths, which were randomly sampled to source the turbulent torque distribution. A wide spectrum of modes is characteristic of MRI as well as of GI, with the main qualitative difference being that MRI has a flatter spectrum than GI or the GI dynamo regime, which are dominated by a few large-scale modes \citep{Deng2020}. Yet since the modes are sampled randomly, as shown by the Gaussian distribution of the resulting turbulent torques, the model, at least qualitatively, is not in conflict with the MRI scenario. Yet in the future, a realistic spectrum tailored to specific turbulence drivers will need to be considered to further assess our quantitative predictions. 

Nevertheless, the detailed properties of MRI in 3D are very complex, showing pattern changes on timescales even longer than the orbital time, and an anisotropic structure of the fluctuating velocity and density field \citep{Murphy2015}. The anisotropy of the turbulent flow, which is expected in the presence of shear but further amplified by the anisotropic nature of the magnetic field, is actually necessary for a net turbulent torque to arise (otherwise force fluctuations will cancel out). Whether or not such  an anisotropic turbulent velocity field results in a turbulent torque that can be characterized by a single gaussian, rather than by a more complex functional for cannot be captured by a simple analytical model. Our simple description of the torque distribution might still hold on long timescales from time-averaging of an anisotropic torque distribution. Ultimately, directly measuring the 3D torques in  MHD simulations is the only viable way to fully account for the physical nature of the flow. Yet, no high-resolution MHD study has reported on the turbulent torques. Our simple proof-of-concept model with a Gaussian distribution for the turbulent torque and other adjustable parameters, such as $\coeff$, is thus, at present, the only way to approach the problem.

{The main focus of this work is to assess how gas turbulence affects thin disk-embedded LISA-observable EMRIs by leaving different imprints on the gravitational waveform than the usually modeled linear dephasing and not how exactly they reach that stage. However, their behavior at a larger scale can dictate the expected rate of such LISA EMRIs. We explore three criteria to assess if our chosen ``golden" EMRI opens a gap: \citep{Crida2006}, \citep{Duffell2013}, and \citep{Gilbaum2025}; Hence, reducing the effectiveness of the stochastic torque due to turbulence. The first two criteria suggest that for our parameter space there is no gap, while \citet{Gilbaum2025} criterion implies that the perturber does opens a gap for $h_0\lesssim0.03$. Since \citet{Gilbaum2025} reference their criterion from low-viscosity ($\alpha\leq0.01$) simulations of \citet{Kanagawa2018}, whose results are a factor of 3 conservative than the similar low-viscosity criterion suggested by a HD study \citet{Duffell2013}. Since we do not consider any $\alpha<0.05$, so its not immediately clear if one can extrapolate these criterion to our parameter space.

But the main thing to consider is that these simulations are carried out for a single perturber on a circular coplanar orbit embedded in a laminar disk.
Due to turbulence, a fluid element could move faster (velocity $v_{\rm turb}$) than the gap-opening timescale, thereby immediately filling the gap. Indeed, \citet{Malik2015} show this for gravito-turbulent disks and \citet{Kubli2026} for MHD disks. For MRI-induced turbulence, which should be most relevant for LISA EMRIs, we provide a simple argument based on MHD simulations. If the largest turbulent eddies are of size $l$ then it is of the same order as the scale height $H$. Hence, the kinematic viscosity is of order $v_{\rm turb}l$ and can be equated to $\alpha c_s H$ (where $c_s$ is the sound speed) to get the relation $v_{\rm turb}\sim\alpha c_s\sim\alpha h_0 \Omega a$. Now, the gap can only open if $v_{\rm turb}$ is much smaller than the migration velocity $v_{\rm lin}$ (i.e, $\dot a_{\rm gas}$ given by type-I torque), since otherwise any gap would be immediately filled due to turbulent flow. If we compare $v_{\rm turb}$ and $v_{\rm lin}$ for $\fedd=1$ at $a=10^4~r_s$ (a typical migration trap location) then we find $v_{\rm turb}>v_{\rm lin}$ for our parameter space. It is to be noted that $v_{\rm turb}$ here is different than the usually computed viscous velocity $v_{\rm vis}$ because of the different characteristic transport length scales: orbital radius for $v_{\rm vis}$ and $H$ for $v_{\rm turb}$; Hence, they relate via $v_{\rm vis}\sim h_0v_{\rm turb}$. This results in qualitative and quantitative differences from \citet{Gilbaum2025} in the gap-opening parameter space. Hence, we believe our results are not severely limited by gap opening at a larger scale.
}

We also do not consider the relativistic corrections to the Lindblad torque \citep{Duque2025a}, which, in principle, can increase the linear torque. Moreover, \citet{Kubli2026} suggests that turbulence can also make the EMRI eccentric, while we only consider circular orbits. The disk structure near EMRIs can also be significantly modified by their epicycle motion with modest eccentricity and inclination \citep{lichenlinwang2022, lichenlin2023}. Since eccentricity is also an early-inspiral effect, it can increase the minimum observable gas dephasing value \citep{Garg2024a,Duque2025b}, and we leave this for future work.

From the waveform modeling perspective, we consider Newtonian-order, spinless quadrupole waveforms. In reality, gas can spin up both component BHs of an EMRI \citep{Bardeen1975,Garg2024c}, and the waveform can include higher modes and beyond-Newtonian corrections \citep{Colpi2024}. However, since we are subtracting the same base waveform when comparing gas-embedded and vacuum EMRIs, ignoring these corrections should be reasonable. Lastly, we use a simple criterion of $8/{\rm SNR}$ to determine whether a given dephasing is observable. In principle, one has to do Bayesian inference to conclude that by varying all parameters of interest \citep{Speri2023,Garg2024b}, since small changes in other parameters can mimic $\dphi$ as a number.

{We note that while convergent migration frequently leads to stable Mean-Motion-Resonances (MMRs) for black holes in laminar disks \citep[e.g.,][]{Epstein-Martin-etal.2025,Moncrieff-etal.2026}, the strong turbulence considered in our chaotic migration model naturally mitigates this complication. Recent hydrodynamical simulations by \cite{Chen2025}, demonstrate that active turbulence significantly enhances resonant overstability, driving migrating bodies to escape from MMRs due to sustained higher equilibrium eccentricities. Consequently, embedded stellar-mass black holes are unlikely to maintain stable resonance chains as they enter the LISA band, rendering mutual MMR perturbations secondary compared to the dominant stochastic gas torques.}

\section{Conclusion}\label{Sec:Conclusion}

We consider the final four-year GW-driven inspiral of the ``golden'' EMRI with source total mass $10^6~\MSun$, $q=5\times10^{-5}$ at $z=0.276$ with ${\rm SNR}=50$, which is embedded in a thin accretion disk with turbulent gas flow, in contrast to the almost always considered laminar disks. The turbulent disk exchanges angular momentum with the binary to apply turbulent torque ($T_{\rm turb}$) on the EMRI instead of the linear torque ($T_{\rm lin}$) expected from the laminar disk, such that its GW phase shifts with respect to the vacuum expectation. This shift manifests as dephasing $\Delta\psi_{\rm turb}$ and will be LISA-observable at the leading-order if it is $\geq{\Delta\psi_{\rm crit}=}8/{\rm SNR}$. We model $T_{\rm turb}$ generally as a Gaussian around $T_{\rm lin}$, inspired by a recent global HD study of gravitational-instability driven disk turbulence, and apply it to our fiducial EMRI, even though it is most likely affected by the magneto-rotation instability. To agnostically consider different sources of turbulence, we introduce two new parameters: $\coeff$ and $N_{\rm max}$ with values $1$ and $\sim100$, respectively, in the aforementioned HD study. While $\coeff$ sets the level of turbulence, $N_{\rm max}$ defines the long-term timescale of expected fluctuations. The resulting turbulent dephasing ($\Delta T_{\rm turb}$) is also a Gaussian around the linear dephasing ($\Delta T_{\rm lin}$). We set $N_{\rm max}=100$ and vary four parameters -- $\fedd$, $\alpha$, $h_0$, and $\coeff$ -- in a reasonable range to bracket the systems of interest where {the realistically maximal} $\Delta\psi_{\rm turb}\geq{\Delta\psi_{\rm crit}}$ but $\Delta\psi_{\rm lin}<{\Delta\psi_{\rm crit}}$. This allows us to report new systems that are thought to exhibit unobservable gas-induced dephasing because of only considering the laminar disk. We find that this happens for $\fedd\gtrsim0.3$, $\coeff\gtrsim{10}$, $h_0\gtrsim0.03$, and $\alpha\gtrsim0.1$ as shown in Fig.~\ref{Fig:Mup2Sig_rad}. We also provide analytical estimates of the mean and standard deviation for $\Delta\psi_{\rm turb}$ in Eq.~\eqref{Eq:analytical_golden}.

Hence, this work strongly motivates MHD simulations of gas disks embedded near-coalescence EMRIs to extract maximum science in the LISA era.

\section*{Acknowledgments}
We acknowledge the very useful discussions with Martin Pessah. We also acknowledge conversations with Barry McKernan, and during GW meetings at the Center for Computational Astrophysics (CCA) at the Flatiron Institute and computational astrophysics meetings at the New York University. {We thank the anonymous referee for their helpful comments that improved this work.} MG acknowledges support from the Swiss National Science Foundation (SNSF) Postdoc Mobility fellowship P500-2\_235363. YW acknowledges the EACOA Fellowship awarded by the East Asia Core Observatories Association. The authors also acknowledge use of the NumPy \citep{harris2020array}, Matplotlib \citep{Hunter2007}, Pandas \citep{Pandas}, and Seaborn \citep{Seaborn} Python packages.

\bibliography{ChaoticEMRIs}
\bibliographystyle{aasjournal}
\normalsize

\appendix

\setcounter{figure}{0} 
\setcounter{table}{0} 
\renewcommand\thefigure{A\arabic{figure}}
\renewcommand\thetable{A\arabic{table}}

\section{Definitions of different variables}
{
In Table~\ref{table:variables}, we summarize different main variables we have introduced throughout the text.}

\begin{table}
\centering
    \begin{tabular}{|>{}C|>{}c|>{}c|}
        \hline
        \pmb{\theta}&{\bf Definition} & {\bf Units}\\
        \hline
        \hline
        M_z &Total redshifted mass & $\MSun$\\
        \hline
        q & Mass ratio& Dimensionless\\
        \hline
        a & Semi-major axis (SMA) & cm\\
        \hline
        \{a_0,a_f\} & Initial and final SMAs & cm\\
        \hline
        r_s & Schwarzschild radius & cm\\
        \hline
        \{\Omega,f\} & Orbital angular and GW frequencies & Hz\\
        \hline
        {\rm SNR} & LISA signal-to-noise ratio of a given EMRI & Dimensionless\\
        \hline
        \alpha& Turbo-viscous coefficient & Dimensionless\\
        \hline
        h_0& Disk aspect ratio & Dimensionless\\
        \hline
        \fedd&Eddington ratio& Dimensionless\\
        \hline
       \Sigma_0& Density normalization defined in Eq.~\eqref{Eq:Sigma} & ${\rm g}\cdot{\rm cm}^{-2}$\\ 
        \hline
        \coeff & Amplitude of turbulence & Dimensionless\\ 
        \hline
        \xi & Torque normalization defined in Eq.~\eqref{Eq:Tgas} & Dimensionless\\ 
        \hline
         \{\mu_\xi,\sigma_\xi\} & Mean and standard deviation of $\xi$ & Dimensionless\\ 
        \hline
        \mathcal{A} & Amplitude of $\dot a_{\rm gas}$ &  Dimensionless\\
        \hline
        v & $(GM_z\pi f/c^3)^{\frac13}$ &  Dimensionless\\
        \hline
        N_{\rm max}& Maximum orbits where $\xi$ is constant & Dimensionless\\
        \hline
        N_{\rm LISA}& Total EMRI orbits in the LISA band & Dimensionless\\
        \hline
        \Delta\psi_{\rm lin} & Gas-induced linear GW dephasing & rad\\
        \hline
        \Delta\psi_{\rm turb} & Gas-induced turbulent GW dephasing & rad\\
        \hline
        \Delta_0 & $\Delta\psi_{\rm turb}$ normalization & rad$\cdot{\rm cm}^{-7}$\\
        \hline
        \{\sigma_{\rm turb},\hat{\sigma}_{\rm turb}\} & Std. deviation of $\Delta\psi_{\rm turb}$ and its analytical approximation & rad\\
        \hline
        \Delta \psi_{\rm crit} & Minimum LISA-observable dephasing, $8/{\rm SNR}$&rad\\
        \hline
        \Lambda_{\rm lin} & $|\Delta\psi_{\rm lin}|/\Delta\psi_{\rm crit}$ & Dimensionless\\
        \hline
        \Lambda_{\rm turb} & $(|\Delta\psi_{\rm lin}|+2\hat\sigma_{\rm turb})/\Delta\psi_{\rm crit}$ & Dimensionless\\
        \hline
    \end{tabular}
\caption{Definitions of the main variables.}
\label{table:variables}
\end{table}

\section{Analytical approximation of the turbulent dephasing Gaussian distribution}\label{App:A}

As given in Eq.~\eqref{Eq:DephasingTotal}, the overall turbulent dephasing is
\begin{align}\label{Eq:DephasingTotal2}
    \Delta\psi_{\rm turb}&= \Delta_0 ~\sum_{j = 0}^K \xi_j \left( a_j^7 - a_{j+1}^7 \right).
\end{align}
The mean of $\Delta \psi_{\rm turb}$ over realizations of the $\{ \xi_i \}$ is 
\begin{align}\label{Eq:mean}
    \langle \Delta \psi_{\rm trub} \rangle_{\xi_i} = \Delta_0~\sum_{j = 0}^K \mu_{\xi} (a_j^7 - a_{j+1}^7) = \Delta_0 \mu_\xi (a_0^7 - a_f^7)=\Delta\psi_{\rm lin},
\end{align}
independent of the realization of the $\{ N_i \}$. Therefore this is also the global average of $\Delta \psi_{\rm turb}$ (over both $\{\xi_i \}$ and $\{ N_i \}$ realizations), and equal to the linear dephasing $\Delta\psi_{\rm lin}$.

We now compute the variance of $\Delta \psi_{\rm turb}$:
\begin{align}
   \sigma^2_{\rm turb} \equiv \langle (\Delta \psi_{\rm turb} - \langle\Delta\psi_{\rm turb}\rangle)^2 \rangle = \Delta_0^2 \left\langle \left(\sum_{j} (\xi_j - \mu_\xi) (a_j^7 - a_{j+1}^7)\right)^2 \right\rangle = \Delta_0^2 \sigma_\xi^2 \sum_j  \langle (a_j^7 - a_{j+1}^7)^2 \rangle,
\end{align}
where we used the independence of the $\xi_j$'s and $a_j$'s, and the fact that $\langle (\xi_i - \mu_\xi)(\xi_j - \mu_\xi) \rangle = \delta_{ij} \sigma_\xi^2$.

We now assume that, over the number of orbital periods $N_i$, the change of SMA is mostly driven by GW emission, and that this change is small enough that we may approximate
\begin{align}
    a_{j+1}&\approx a_j (1-N_j ~g(a_j)), \label{eq:ajp1-aj}\\
    {\rm where}~~g(a_j)&\approx-\frac{\dot{a}_{\rm GW}(a_j)}{a_j}\frac{2\pi}{\Omega}=\frac{16\sqrt{2}\pi}{5}q~(r_s/a_j)^{\frac52} = g(a_0) (a_0/a_j)^{5/2}\nonumber.
\end{align}
This should be a fair approximation sufficiently far away from merger. Hence we have
\begin{align}
    (a_j^7 - a_{j+1}^7)^2 \approx a_j^{14} \left( 1- (1 - N_j g(a_j))^7\right)^2 \approx 49~ a_j^{14} g^2(a_j) N_j^2 = 49~ g^2(a_0)a_0^5 ~ a_j^9 N_j^2.
\end{align}
Hence the variance of analytical turbulent dephasing becomes
\begin{align}
    \hat{\sigma}^2_{\rm turb} \approx 49~ \Delta_0^2~\sigma_\xi^2~g^2(a_0) a_0^5~\sum_j \langle N_j^2 a_j^9 \rangle. \label{eq:sigma_turb2}
\end{align}
Using Eq.~\eqref{eq:ajp1-aj} recursively, we further obtain $a_j \approx a_0 \prod_{l = 0}^{j-1} (1 - N_l g(a_l))$. Since this expression is only valid for small $N_l g(a_l)$, we may substitute $a_l \approx a_0$. Physically, this translates to the fact that most of the dephasing is coming from the initial evolution. We thus find
\begin{align}
    \langle N_j^2 a_j^9\rangle \approx a_0^9 \langle N_j^2 \prod_{l=0}^{j-1} (1 - g(a_0) N_l)^9\rangle = a_0^9 \langle N_j^2 \rangle \prod_{l = 0}^{j-1} \langle (1 - g(a_0) N_l)^9 \rangle \approx a_0^9 \langle N^2 \rangle (1 - 9 g(a_0) \langle N \rangle)^j,
\end{align}
where we used the fact that different $N_j$ are independent. Inserting this result into Eq.~\eqref{eq:sigma_turb2}, we finally arrive at
\begin{align}
    \hat{\sigma}_{\rm turb}^2 \approx 49 ~\Delta_0^2 \sigma_\xi^2 g^2(a_0) a_0^{14} \langle N^2 \rangle \frac{1 - (1 - 9 g(a_0) \langle N \rangle)^K}{9 g(a_0) \langle N \rangle} \approx \frac{49}{9} \Delta_0^2 \sigma_\xi^2 g(a_0) a_0^{14} \frac{\langle N^2 \rangle}{\langle N \rangle},
\end{align}
where we assumed $K \gg 1$ to approximate the numerator by 1. Since $N$ is uniformly distributed between 0 and $N_{\rm max}$, we get $\langle N^2 \rangle = N_{\rm max}^2/3$ and $\langle N \rangle = N_{\rm max}/2$, implying
\begin{align}
    \hat{\sigma}_{\rm turb} \approx \frac{5}{17} \sqrt{\frac{5\pi}{3\sqrt{2}}} \frac{G^2M_z}{c^4q^{\frac52}}\left(\frac{a_0}{r_s}\right)^{\frac{23}{4}} \sigma_\xi \sqrt{N_{\rm max}}.
\end{align}
\renewcommand\thefigure{C\arabic{figure}} 
\section{Viscous disk profile}\label{App:Sigma_vis}
{
Since \citet{Wu2024} considers a viscous profile $\Sigma_{\rm vis}\propto a^{-1/2}$, we also compute some important results for it here.

In physical units, the surface density can be expressed as
\begin{equation}
    \Sigma_{\rm vis}=\frac{2.37\times10^5}{{\rm g}\cdot{\rm cm}^{-2}}\left(\frac{a}{r_{\rm s}}\right)^{-\frac12}\left(\frac{\fedd}{1.0}\right)\left(\frac{\alpha}{0.1}\right)^{-1}\left(\frac{h_0}{0.03}\right)^{-2}
\end{equation}

We get the above disk profile by assuming quasi-steady state accretion and fixing $\{\fedd,h_0,\alpha\}$:
\begin{align}
    \dot M=\fedd\dot{\rm M}_{\rm Edd}&=\fedd\frac{M}{50~\rm{Myr}}\approx3\pi\nu(a)\Sigma(a)=3\pi\alpha\Omega h^2_0a^2\Sigma(a),\\
    \implies \Sigma(a)&\approx\frac{\fedd}{3\pi \times 50~{\rm Myr}}\frac{M}{\alpha h^2_0\sqrt{GM}}r^{-\frac12}=\frac{\fedd}{3\pi \times 50~{\rm Myr}}\frac{c}{\alpha h^2_0G\sqrt{2}}\left(\frac{a}{r_s}\right)^{-\frac12},\nonumber\\
    \implies \Sigma(a)&\approx2.37\times10^5 \left(\frac{\fedd}{1.0}\right)\left(\frac{\alpha}{0.1}\right)^{-1}\left(\frac{h_0}{0.03}\right)^{-2}\left(\frac{a}{r_s}\right)^{-\frac12},\nonumber
\end{align}

Then similar to the computations for the radiation-dominated disk in Eq.~\ref{Eq:analytical_golden}, we can directly define the linear torque and the analytical turbulent dephasings for the viscous disk which embeds the golden EMRI:

\begin{align}
    \Delta\psi^{\rm(golden)}_{\rm lin,vis}&=-3.85\left(\frac{\fedd}{1.0}\right)\left(\frac{\alpha}{0.1}\right)^{-1}\left(\frac{h_0}{0.03}\right)^{-{4}}~{\rm rad},\\
    \hat\sigma^{\rm(golden)}_{\rm turb,vis}&={7.13}\times10^{-2}\left(\frac{\alpha}{0.1}\right)^{\frac12}\left(\frac{h_0}{0.03}\right)\left(\frac{\coeff}{1.0}\right)\left(\frac{N_{\rm max}}{100}\right)^{\frac12}~{\rm rad},
\end{align}
where the linear torque here has the pre-factor $-1.11$ \citep{Jimenez2017}. 

In Fig.~\ref{Fig:Mup2Sig_vis}, we show the effect of the turbulent torque for the viscous disk only for $\coeff=1$. Even for low turbulence amplitude, the maximal turbulent dephasing is observable for all parameters.

\begin{figure*}
    \centering
    \includegraphics[width=\linewidth]{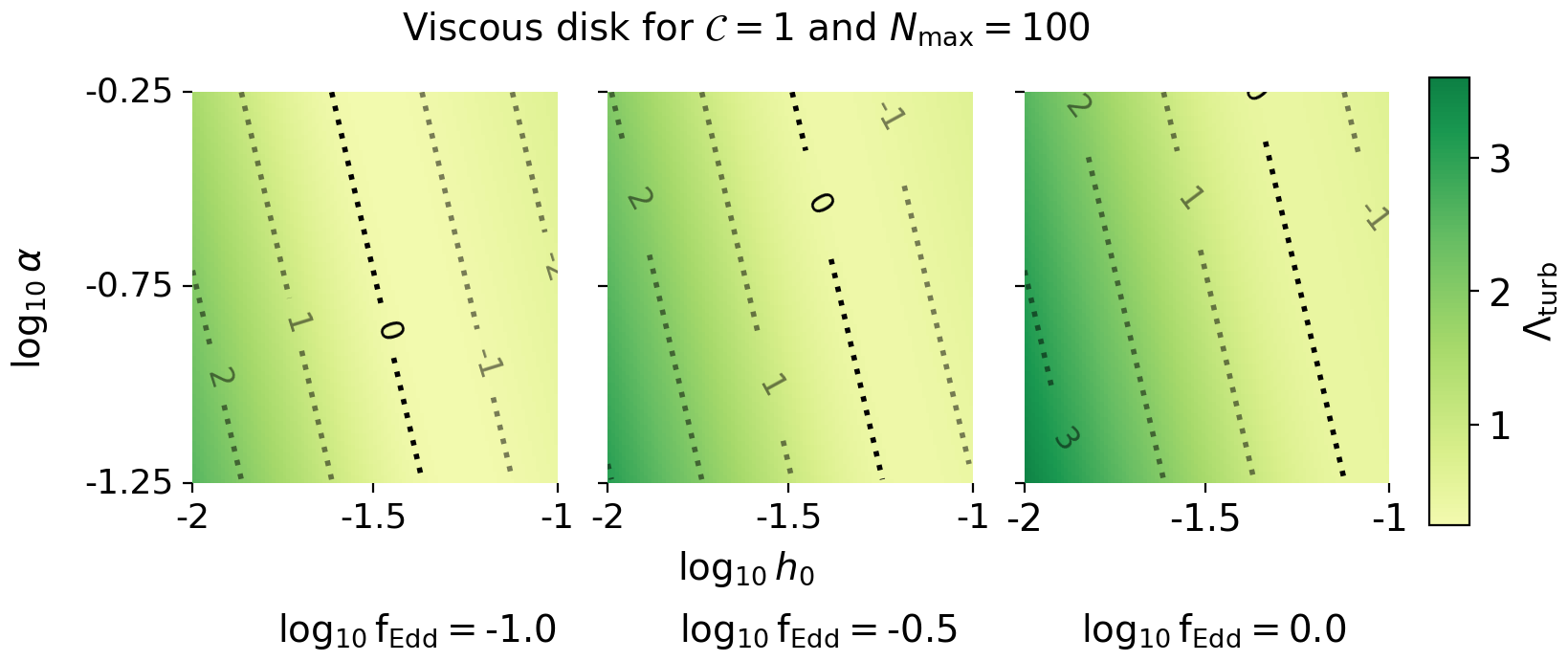}
    \caption{Same as Fig.~\ref{Fig:Mup2Sig_rad} but for the viscous disk and only the $\coeff=1$ row. Gas-induced dephasings across the parameter space could become LISA-observable.}
    \label{Fig:Mup2Sig_vis}
\end{figure*}

}

\section{Error on $\hat{\sigma}_{\rm turb}$}\label{App:B}
\renewcommand\thefigure{D\arabic{figure}}
\setcounter{figure}{0}  

To compare the analytical approximation of $\hat{\sigma}_{\rm turb}$ in Eq.~\eqref{Eq:analytical_golden} to the numerical standard deviation $\tilde{\sigma}_{\rm turb}$, we do $20000$ realizations of $\Delta\psi_{\rm turb}$ using Eq.~\eqref{Eq:DephasingTotal}  by considering these parameter combinations {for $N_{\rm max}=100$ and $\fedd=1$ for the radiation-dominated disk:}
\begin{itemize}\label{Eq:params2}
    \item $\coeff\in{\{1,10,100\}}$,
    \item $\alpha\in\{10^{-1.25},10^{-1},10^{-0.75},10^{-0.5},10^{-0.25}\}$,
    \item $h_0\in\{10^{-2},10^{-1.75},10^{-1.5},10^{-1.25},10^{-1}\}$.
\end{itemize}

For this purpose, we construct a grid for the SMA over $N_{\rm LISA}$ orbits in Eq.~\eqref{Eq:golden}, such that
\begin{equation}
    a_{j+1}\approx a_j(1-g(a_j)).
\end{equation}
Since dephasing accumulates mainly at the initial separation, the above approximation should be fine.

\begin{figure*}
    \centering
    \includegraphics[width=\linewidth]{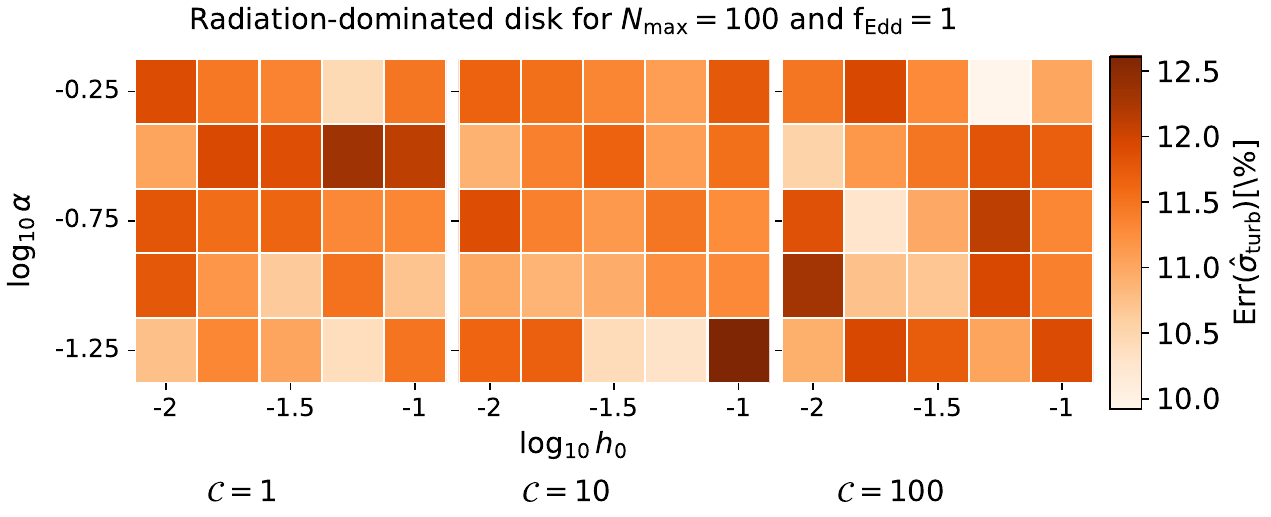}
    \caption{Percentage error ${\rm Err}(\hat{\sigma}_{\rm turb})[\%]$ on the analytical standard deviation $\hat{\sigma}_{\rm turb}$ of the turbulent dephasing distribution with respect to the numerical standard deviation $\tilde{\sigma}_{\rm turb}$ estimated from $20000$ realizations. {We show errors only for $N_{\rm max}=100$ and $\fedd=1$, vary $\coeff\in{\{1,10,100\}}$ and} have a coarse $5\times5$ grid for each $\{\alpha,h_0\}$ panel. The figure shows that ${\rm Err}(\hat{\sigma}_{\rm turb})[\%]\sim10$ across the parameter space.}
    \label{fig:Sigma_error}
\end{figure*}

Finally, we calculate the percentage error on $\hat{\sigma}_{\rm turb}$ with respect to the numerical result:
\begin{equation}
    {\rm Err}(\hat{\sigma}_{\rm turb})[\%]=100\times\frac{(\hat{\sigma}_{\rm turb}-\tilde{\sigma}_{\rm turb})}{\hat{\sigma}_{\rm turb}},
\end{equation}
and show it in Fig.~\ref{fig:Sigma_error}. Across the parameter space {(applicable also to varying $\fedd$)}, the error remains close to $10\%$, indicating that $\hat{\sigma}_{\rm turb}$ in Eq.~\eqref{Eq:analytical} is a good approximation.

\end{CJK*}
\end{document}